# Pixelated Bayer Spectral Router Based on Sparse Meta-atom Array

**Yifan Shao, Rui Chen, Yubo Wang, Shuhan Guo, Junjie Zhan, Pankaj K. Choudhury, and Yungui Ma***

State Key Lab of Modern Optical Instrumentation, Centre for Optical and Electromagnetic Research, College of Optical Science and Engineering; International Research Center (Haining) for Advanced Photonics, Zhejiang University, Hangzhou, 310058, China

**Corresponding author's E-mail:** yungui@zju.edu.cn

**Abstract**

It has long been a challenging task to improve the light collection efficiency of conventional image sensors built with color filters that inevitably cause the energy loss of out-of-band photons. Although various schemes have been proposed to address the issue, it is still very hard to make a reasonable tradeoff between device performance and practicability. In this work, we demonstrate a pixelated spectral router based on sparse meta-atom array, which can efficiently separate the incident R (600-700 nm), G (500-600 nm), and B (400-500 nm) band light to the corresponding pixels of a Bayer image sensor, providing over 56% signal enhancement above the traditional color filter scheme. The CMOS-compatible spectral router has superior characteristics of polarization insensitivity and high incident angle tolerance (over 30°), enabled by simple compound $Si_3N_4$ nanostructures which are very suitable for massive production. Imaging experiments are conducted to verify its potential for real applications. Our pixelated spectral router scheme is also found to be robust and could be freely adapted to image sensors of various pixel sizes, having great potential in building the new generation of high-performance image sensing components.

## 1. Introduction

With the rapid development of smartphones and digital cameras, the pixel size of image sensors is constantly shrinking to meet the requirement of high-resolution imaging.[1] The reduction in pixel size will diminish the sensitivity, signal-to-noise ratio and dynamic range of image sensors due to less light energy received by each pixel. In order to increase the photosensitive area, image sensors have evolved from front-illuminated to back-illuminated[2] and then to stacked configurations.[3] However, image sensors typically rely on spectral filter arrays embedded above photodetectors to extract color information, only transmitting certain narrow-band light while filtering out the rest.[4] The out-of-band photon loss seriously limits the energy utilization efficiency of image sensors and inevitably degrades the imaging performance, especially for high-density imaging in low-light scenarios. Nanophotonic spectral filters[5-8] have been developed to replace traditional pigment color filters,[9] but they still cannot fundamentally address the issue of out-of-band energy waste. As for Bayer image sensors, even if the color filters reach 100% transmittance in the passband, the maximum energy utilization efficiencies of red (R), green (G), and blue (B) channels are still limited to only 25%, 50%, and 25%, respectively.[4]

To tackle this problem, spectral splitting techniques that aim to spatially separate light of different bands to corresponding photodetectors have been investigated. Similar to spectrometers, blazed gratings and microlens array were early combined to converge light of different wavelengths onto distinct pixels.[10] To avoid the issue of inadequate angular dispersion in the same diffraction order of blazed gratings, color separation gratings that can divide R, G, and B bands to +1, 0, and -1 order directions, respectively, were proposed.[11,12] However, they work for the far-field regimes and usually require long propagation distances and large grating periods which limit the pixel-level integration with photodetector arrays. Many plasmonic nanoantennas capable of achieving wavelength-dependent directional scattering have been studied, but their efficiencies are constrained by the inherent loss of metals.[13,14] To avert ohmic loss, dielectric scatterers and diffractive optical elements were utilized for spectral routing.[15-19] Nonetheless, due to severe crosstalk between distinct color channels, most of these designs rely on a post-processing reconstruction algorithm (e.g., conversion matrix method) to restore the RGB color information, which will have a poor effect for imaging dark scenarios.[12,15,17] Additionally, researchers have designed micro-metalens arrays to focus light at different wavelengths onto specific pixels.[20-25] Some of these designs built a library containing a variety of intricate unit cells in order to approximate routing phase profiles of multiple wavelengths simultaneously,[20,21] while the others sacrificed a certain amount of efficiency because of the spatial multiplexing method.[24,25] Recently, spectral routers based on code-like or freeform metasurfaces were reported via inverse design.[26-30] Moreover, inverse design algorithms can be further applied to develop high-performance spectral routers utilizing three-dimensional (3D) metamaterials.[31-40] However, their topologies are too complicated to be feasible for the visible light application even by the state-of-the-art processing technologies.

Herein, we demonstrate a highly efficient pixelated Bayer spectral router that can precisely separate and focus the incident R (600-700 nm), G (500-600 nm), and B (400-500 nm) light to the corresponding pixels of a typical RGGB pattern image sensor, based on a metasurface made of the array of sparse meta-atoms (i.e., unit cells larger than the operating wavelength). We prove that the pixel-level dispersion engineering could efficiently split the broadband photons by relatively simple supercells consisting of only four isolated square nanopillars, meeting an excellent balance between structural complexity and device performance. As a result, high average spectral routing efficiencies of 51.13%, 62.91%, and 42.57% for R, G, and B bands, respectively, are obtained, corresponding to a 56.6% signal enhancement beyond the classic Bayer color filter scheme. The spectral router's applicability in color imaging is experimentally validated, obviating the need for any complex reconstruction algorithm due to the eliminated crosstalk resulting from its combination with color filters. The proposed device also exhibits good polarization insensitivity by employing the diagonally symmetric structural layout and meta-atoms with high rotational symmetries. We also show that the response of the router under oblique incidence could be enhanced by introducing a structure shift method, which can expand the maximum acceptable chief ray angle to over 30°. The structural layout of the device is very straightforward with a minimum feature size larger than 150 nm, and all gaps between nanostructures are larger than 500 nm, which makes it suitable for massive industrial production through conventional nanofabrication technologies, such as deep ultraviolet (DUV) lithography[41] and nanoimprinting.[42,43] The pixelated Bayer spectral router developed here is believed to be of practical significance for building the next generation of high-performance image sensors.

## 2. Results

### 2.1. Principle and Design of the Bayer Spectral Router

Figure 1a schematically illustrates a conventional image sensor consisting of the microlens array, color filters, and photodetectors. The most common color filter array arrangement for image sensors in the visible band is the Bayer pattern (RGGB), with each Bayer cell containing one R, two G, and one B color filters. Due to the selective absorption of color filters, a great deal of energy is wasted, thereby limiting the maximum detecting efficiency to 33.33%.[4] The goal of this work is to design a pixelated spectral router which can improve the energy utilization efficiency of image sensors. Figure 1b schematically illustrates an image sensor with the spectral router, which is capable of efficiently routing the R, G, and B band light onto the corresponding pixels of the Bayer pattern. The color filter array is also configured between the spectral router and photodetectors to filter the routed light to reduce the crosstalk between different color channels.

To manipulate the light field within small apertures, sophisticated nanostructures with complicated distributions have been previously employed to satisfy specific dispersive phase profiles.[21,29] In this work, to facilitate massive production, we aim to explore the feasibility of realizing color separation at pixel level-dimensions through simple nanostructures made of few regular-shaped sparse meta-atoms while ensuring high performance. Figure 1c,d illustrate the structural layout of the spectral router, i.e., one supercell of the metasurface. The supercell has a period of 2.24 μm × 2.24 μm, corresponding to four pixels in one Bayer cell (RGGB) of the image sensor. Therefore, the spectral router is well matched to image sensors with the pixel size of 1.12 μm × 1.12 μm, which is a common pixel size used in current commercial complementary metal-oxide-semiconductor (CMOS) image sensors. In this design, each meta-atom includes one $Si_3N_4$ square nanopillar above the $SiO_2$ substrate. $Si_3N_4$ is almost transparent in the visible band and compatible with the CMOS processing technology,[44] which is very beneficial to the enhancement of device efficiency and the integration with existing image sensors. The refractive index parameters of $Si_3N_4$ and $SiO_2$ used here are plotted in Supporting Information Figure S1. There are four meta-atoms in one supercell, with each grid centered by a $Si_3N_4$ nanopillar. In contrast to the routers made up of subwavelength unit cells, the sizes of sparse meta-atoms used in our design can be larger than the operating wavelength. The transmission phase profile of the supercell will be modulated by $Si_3N_4$ nanopillars and also the air gaps between them (or corresponding to a gap phase).[45] The performance of the spectral router is optimized by iteratively adjusting the structures of meta-atoms. The particle swarm optimization (PSO) algorithm,[46] which is an intelligent stochastic search technique for finding the optimal solution, has been combined with the finite-difference time-domain (FDTD) numerical simulation to optimize design parameters, aiming to achieve high spectral routing efficiency. The method and settings of the electromagnetic simulation are elucidated in Methods Section. The optimization process of the spectral router is shown schematically in Supporting Information Figure S2. The design parameters include the widths of $Si_3N_4$ nanopillars ($w_1$, $w_2$, and $w_3$), the height of $Si_3N_4$ nanopillars ($h$), and the distance between nanopillars and the detecting plane ($h_d$). Due to the diagonal symmetry of the Bayer pattern (RGGB) and to ensure the polarization insensitivity of the device, nanopillars on the two G pixels have the same width ($w_2$). To endow the spectral router with the potential for massive production in industrial applications, a constraint condition that widths of all square nanopillars exceed 150 nm is imposed during the optimization process. To ensure the compactness and reduce crosstalk between adjacent supercells, we set $h_d$ to be within 4.5 μm. The spectral routing efficiency at a specific wavelength is defined as the ratio of the light energy reaching the pixels of the corresponding band to the total energy incident on the supercell at this wavelength. In order to achieve high routing performance across broadband, the average spectral routing efficiencies of R, G, and B bands serve as crucial indicators in evaluating the device performance, which are defined as

$$\text{Eff}_R = \frac{1}{\Delta \lambda_R} \int_{600 \text{ nm}}^{700 \text{ nm}} T_R(\lambda) \, d\lambda, \tag{1}$$

$$\text{Eff}_G = \frac{1}{\Delta \lambda_G} \int_{500 \text{ nm}}^{600 \text{ nm}} T_G(\lambda) \, d\lambda, \tag{2}$$

$$\text{Eff}_B = \frac{1}{\Delta \lambda_B} \int_{400 \text{ nm}}^{500 \text{ nm}} T_B(\lambda) \, d\lambda, \tag{3}$$

respectively, where $T_R(\lambda)$, $T_G(\lambda)$, and $T_B(\lambda)$ are spectral routing efficiencies of R, G, and B channels, respectively, as a function of wavelength, while $\Delta\lambda_R = \Delta\lambda_G = \Delta\lambda_B = 100$ nm are bandwidths of R, G, and B bands. The efficiency enhancement factor is defined as

$$\text{Eff} = \text{Eff}_R + \text{Eff}_G + \text{Eff}_B. \tag{4}$$

It means that the efficiency enhancement factor of an ideal conventional Bayer filter array is Eff = 25% + 50% +25% = 1. Intending to increase the spectral routing efficiency of the device, the efficiency enhancement factor is set as the figure of merit (FOM) in the optimization.

After sufficient optimization, the ultimate design parameters are determined as follows: $w_1$ = 920 nm, $w_2$ = 150 nm, $w_3$ = 278 nm, $h$ = 998 nm and $h_d$ = 4 μm. Figure 2a draws the transmission phase profiles of one optimized supercell at three typical wavelengths: 650 nm, 550 nm, and 450 nm. These phase distributions are very close to those of the ideal microlens shown in Figure 2b, especially for the R and B light. Therefore, the sparse meta-atoms collectively function as a pixel-level dispersion-engineered micro-metalens, which can separate and focus light of different colors to corresponding pixels. In Supporting Information Section S3, we simulate the focusing effect for R, G, and B color under limited aperture, where the incident light only cover nine meta-atoms. For each R/G/B pixel, the nine meta-atoms corresponding to the central pixel and the eight surrounding pixels function as a microlens, focusing light towards the central pixel. In this manner, the central pixel collects the energy that would otherwise be absorbed by color filters of the neighboring pixels of different colors, thereby improving the energy utilization efficiency. Additionally, the simulation results demonstrate that the contribution to the focusing effect primarily comes from the central meta-atoms and the adjacent meta-atoms on the pixels of colors different from the central pixel, indicating that the spatial crosstalk of the same color channel between supercells is small. For the sparse meta-atom array, Figure 2c-e illustrate the simulated power flow density distributions of R (650 nm), G (550 nm), and B (450 nm) light on the detecting plane (Z = -4 μm, i.e., 4 μm below the nanopillars), respectively, which are already the average results under the X- and Y-axis linear polarization incidence conditions. Obviously, most of the R, G, and B light is routed to their targeted pixels. In order to demonstrate the phenomenon of wavelength-dependent routing in the propagation direction intuitively, Figure 2f-h present the power flow density distributions of the XZ cross section at wavelengths of 650, 550 and 450 nm, respectively. The coordinate system is shown in Figure 1c. The Y coordinates of the XZ cross section are 0.28 μm for Figure 2d and -0.56 μm for Figure 2e,f. Figure 2g plots the simulated spectral routing efficiencies of R, G, and B channels. The horizontal dashed lines in Figure 2i represent the maximum spectral routing efficiencies of an ideal traditional Bayer color filter array. The area above the dashed lines and below the spectral routing efficiency curves is the total enhanced energy, which contributes to higher efficiency enhancement factor. The peak spectral routing efficiencies of the spectral router are 54.62%, 70.43% and 64.3% for R, G, and B light. The average spectral routing efficiencies are 51.13%, 62.91%, and 42.57% for R, G, and B bands, respectively. Therefore, the efficiency enhancement factor is 1.566, which means that 56.6% signal enhancement is obtained by the spectral router.

## 2.2. Experimental Demonstration of Pixelated RGB Separation

In order to experimentally verify the RGB separating function of the designed device, we have fabricated the metasurface by plasma-enhanced chemical vapor deposition (PECVD), electron beam lithography (EBL), lift-off and reactive-ion etching (RIE). Details of the fabrication process are presented in Methods Section and Supporting Information Figure S4. The top-view and tilted-view scanning electron microscopy (SEM) images of the sample are presented in Figure 3b,c. After completing the fabrication, the spectral routing effect of the device is characterized using the optical measurement setup shown in Figure 3a. Color filters can be selectively inserted between the collimation system and the spectral router according to the wavelength or band required to be measured. First of all, the spectral router is illuminated by the collimated beam of white light-emitting diode (LED, Thorlabs MNWHL4, 4900K). As shown in Figure 3d, the image on the detecting plane of the spectral router is magnified by the microscope and captured by a color image sensor (HW200). Figure 3e-g present the measured intensity profiles of R, G, and B channels on the detecting plane, respectively. Agreeing with the simulation results, it can be observed that R, G, and B light is effectively routed to the corresponding pixels of the Bayer pattern. Supporting Information Figure S5 presents the intensity profiles on the detecting plane under the illumination of light at different wavelengths (step: 20 nm), which are measured by the same experimental setup with a monochromic image sensor (MV-CS200-10UM). Figure 3h plots the measured spectral routing efficiencies of R, G, and B channels. The deviations between the simulation and experimental results mainly arise from the processing errors of the sample, collimation of the light source and diffraction limit of the microscope system. The measured average spectral routing efficiencies of the router are 49.53%, 55.27%, and 37.74% for R, G, and B bands, respectively, which greatly exceed the spectral routing efficiencies of traditional Bayer color filter array.

## 2.3. Color Imaging with the Bayer Spectral Router

We experimentally verify the applicability of the spectral router in color imaging by the optical setup shown in Figure 4a. An object is placed at a certain distance (665 mm for the ColorChecker image and 540 mm for the Rubik's Cube image) away from the imaging system and is imaged onto the spectral router by an imaging lens (F#6, focal length $f$ = 5 mm). The microscope, R/G/B bandpass filters and a monochromic image sensor (MV-CS200-10UM) are combined to mimic the architecture of an image sensor with color filters and photodetectors. Supporting Information Figure S6 depicts the spectra of R/G/B bandpass color filters. Because R, G, and B color filters are employed in sequence to imitate the Bayer color filter array on the detecting plane to eliminate crosstalk between channels, there is no need to use any complex algorithm (e.g., conversion matrix method) during the image reconstruction process, which is unfavorable in dark scenarios due to strong noise. In the case where the object is an image of ColorChecker, Figure 4b-d present the measured intensity profiles of R, G, and B channels, respectively, on the detecting plane after the image is routed. Then, the mosaic images of R, G, and B channels can be obtained directly according to the calibrated spectral routing efficiency. As shown in Supporting Information Figure S7d-g, after demosaicing through bilinear interpolation,[47] the images of R, G, and B channels are reconstructed respectively. Figure 4e presents the reconstructed color image after performing white balance.[48] As a contrast, as shown in Supporting Information Figure S7h-k, the reference images of R, G, and B channels obtained using only color filters without the spectral router are also reconstructed through color correction and demosaicing. Obviously, the reconstructed images with the spectral router are in good agreement with the reference images for all color channels. Figure 4f,g present the reference color image obtained with only color filters, which is consistent with Figure 4e. The intensity of Figure 4f is matched to that of Figure 4e for clearer comparison. Figure 4h-m present the experimental results analogous to Figure 4b-g, reconstructing the color image of Rubik's Cube. There is a small speckle in the lower left corner of the reconstructed image of Rubik's Cube due to the fabrication defects at that position. More details of the reconstruction of images of ColorChecker and Rubik's Cube are provided in Supporting Information Section S7 and Section S8, respectively. Supporting Information Figure S9 shows the efficiency enhancement for color imaging with the Bayer spectral router. The average efficiency enhancement factor is 1.3 compared with the traditional color filter scheme. This value is comparatively lower than the case without an imaging lens (normal incidence). Note that here we apply the combined scheme (spectral router + color filters) to enhance the detecting efficiency without a post matrix conversion. If only the spectral router is utilized together with the conversion matrix method,[12,15,17] the energy utilization efficiency can be enhanced by approximately 3 times compared with the conventional filter scheme, as all transmitted energy of the router is utilized. But this approach will suffer from crosstalk between color channels. Additionally, limited by the finite field of view area caused by the microscopy system, the image of ColorCheck and Rubik's Cube only occupies $132 \times 88$ and $110 \times 110$ pixels, respectively. The numerical aperture and aberrations of the imaging system also blur the images. The spectral router is anticipated to be directly integrated with image sensors in the future, enabling its application in clear, large-area, high-resolution imaging.

## 3. Discussion

To further explore the potential of this Bayer spectral router for practical industrial applications, we discuss its features of polarization response, incident angle tolerance and fabrication feasibility. The spectral routing efficiencies of R, G, and B channels under the illumination of different polarization states are shown in Supporting Information Figure S10. Since the structural layout of the supercell is diagonally symmetric and all $Si_3N_4$ nanopillars have four-fold symmetric cross sections, the spectral router is insensitive to polarization, which is applicable to imaging scenarios with arbitrary polarized illumination. Considering the numerical aperture (NA) of practical imaging systems, the light incident on image sensors contains a certain range of angles.[49] Therefore, the tolerance to incident angle is a significant parameter of the spectral router. Supporting Information Figure S11a-c present the numerically calculated spectral routing efficiencies of R, G, and B channels under different incident angles. The average efficiency of R, G, and B channels remains above 33.33% up to a 10° incident angle, which is higher than the efficiency of an ideal Bayer color filter array. According to the same definition (the angle where the efficiency drops to half of the maximum) used in previous studies,[20,31,36] the maximum acceptable incident angle for this spectral router is 14.6°, corresponding to the NA of 0.252 for the imaging lens.

Under oblique incidence of large angles, as shown in Supporting Information Figure S11d-f, a portion of light intended to be routed to the $R/G_1/G_2/B$ pixels is misrouted to other neighboring pixels, leading to the decreased spectral routing efficiency and the aggravation of crosstalk between channels. Referring to the mature microlens design of edge pixels in conventional image sensors,[50] as shown in Figure 5a, we employ the structure shift method to improve the device performance under oblique incidence. The arrows are just a view guidance for the diffraction of incident light. Under the incident angle of $\theta$, the displacement of nanostructures is calculated by

$$\Delta x(\theta) = h \times \tan\left[\arcsin\left(\frac{\sin\theta}{n_{\text{eff}}}\right)\right] + h_d \times \tan\left[\arcsin\left(\frac{\sin\theta}{n_{SiO_2}}\right)\right], \quad (5)$$

where the former term represents the required shift caused by the nanostructure layer and the latter term represents the required shift caused by the $SiO_2$ spacer layer, $n_{\text{eff}}$ is the effective refractive index of the nanostructure layer[51] and $n_{SiO_2}$ is the refractive index of $SiO_2$. Approximately, most of the misrouted light can be transferred to the desired pixels by shifting the structures according to the chief ray angle corresponding to the pixel position. For example, Figure 5d-f illustrate the simulated power flow density distributions of R, G, and B channels under 10° incidence, exhibiting the improved performance by shifting structures ($\Delta x$ = 626 nm). As shown in Figure 5g-i, by employing the structure shift method, the spectral routing efficiencies of R, G, and B channels remain nearly constant when the incident angle increases to 10°. Figure 5b plots the efficiency enhancement factor as a function of the incident angle with and without shifting structures, with the black horizontal dashed line representing the efficiency enhancement factor of an ideal conventional Bayer filter array. It intuitively reflects the significant role of the structure shift method in improving efficiency under oblique incidence. In the case of no shifting, the efficiency enhancement factor increases instead when the incident angle is larger than 18.5° because the obliquely traveling photons have reached the pixels of corresponding bands of the next supercell. Figure 5c plots average spectral routing efficiencies of R, G, and B channels as a function of the incident angle when using the structure shift method. The purple and green horizontal dashed lines represent the maximum efficiencies of an ideal Bayer color filter array for R/B (25%) and G (50%), respectively. Simulation results reveal that the structure shift method can expand the maximum acceptable chief ray angle to over 30°, which is applicable to image sensors with large areas. As shown in Supporting Information Figure S11h, the spectral routing efficiency under oblique incidence can even be further improved by specially optimizing the nanostructures while keeping the height of nanopillars and the position of detecting plane unchanged ($h$ = 998 nm, $h_d$ = 4μm).

Compared with previous pixelated light splitting techniques based on micro-metalens arrays[20,21], code-like and freeform metasurfaces,[26-29] our spectral router based on sparse meta-atoms has a straightforward structural layout comprising only four nanopillars in each supercell, with a large feature size of 150 nm and all gaps between nanopillars larger than 500 nm, thus suitable for massive production through mature processing technologies, such as deep ultraviolet (DUV) lithography[41] and nanoimprinting.[42,43] The detailed comparison between the structural layouts of this work and those code-like metasurfaces is illustrated in Supporting Information Figure S12. Moreover, our spectral router notably improves the spectral routing efficiency of the G channel while maintaining high performance of the R and B channels. As shown in Table S1 of Supporting Information Section S12, compared with past researches, our spectral router based on the array of sparse meta-atoms simultaneously realizes high spectral routing performance and fabrication simplicity. Practically, we use sparse meta-atoms with fewer variables to realize similar or even better performance compared to previous spectral routers that have sophisticated nanostructures. Using similar patterns, in Supporting Information Section S13, we design two Bayer spectral routers that can match the pixel sizes of 1 μm × 1 μm and 0.5 μm × 0.5 μm, respectively, while the latter belongs to the state-of-the-art image sensors with the smallest pixel size.[52] Both show very good pixelated color splitting performance. These results indicate our sparse meta-atoms based spectral router scheme is robust and applicable to image sensors of high-density pixels. Furthermore, in practical applications, the color filter array can be integrated between the spectral router and photodetector array to improve color purity. As shown in Supporting Information Section S14, we simulate the architecture including the spectral router, $SiO_2$ spacer layer, color filters, anti-reflection film and photodetectors from top to bottom. It is revealed that the crosstalk between different color channels can be significantly suppressed by the color filter array, which is highly beneficial for improving the performance of image sensors.

## 4. Conclusion

In conclusion, we have demonstrated a pixelated spectral router based on sparse meta-atom array and experimentally verified its potential in color imaging for true scenarios. This optical hardware can improve the photon collection efficiency of CMOS image sensors by a ratio of 56.6% above the classic color filter scheme. The spectral routing efficiency is insensitive to the incident light polarization and keeps high values at large incident angles realized by flexibly shifting the nanostructures. Compared with the existing blueprints, the simple structural layout and CMOS compatibility processing techniques are the most important features of our design, enabling massive production for industrial applications. Putting these into perspectives, it is highly believed that pixelated spectral routers based on sparse meta-atoms can play a key role in developing the new generation of advanced image sensors.

## 5. Methods

**Numerical Simulation:** In this work, the full-wave electromagnetic simulation of this Bayer spectral router is executed by the commercial software Lumerical FDTD Solutions. We first model the supercell of the spectral router which includes four $Si_3N_4$ nanopillars on a $SiO_2$ substrate. The period of the supercell is 2.24 μm × 2.24 μm. Design parameters can be assigned to the model in real time to adjust structures during the optimization. Supporting Information Figure S1 plots the refractive indexes of used materials. The plane wave source is incident along the -Z direction. The periodic boundary condition is employed in both X and Y directions. The perfect matched layer (PML) boundary condition is employed in the Z direction. The mesh size of the simulation is set to 20 nm. Four frequency domain field and power monitors with a pixel size of 1.12 μm × 1.12 μm are arranged $h_d$ away from the $Si_3N_4$ nanopillars to obtain the power flow distributions on the detecting plane, which can be used to evaluate the performance of the device and calculate the spectral routing efficiency. As shown in Supporting Information Section S2, the PSO algorithm is combined with the FDTD simulation to optimize design parameters of the spectral router to achieve high efficiency enhancement factor.

**Sample Processing:** The flow diagram of the sample processing is shown in Supporting Information Figure S4. The quartz substrate is cleaned in the ultrasonic bath of acetone, isopropyl alcohol (IPA), and deionized water for 10 minutes, respectively. A $Si_3N_4$ film with a thickness of 998 nm is deposited on the substrate by PECVD. Then, a 170 nm thick positive electron-beam photoresist (ZEP520A) is spin-coated on the $Si_3N_4$ layer. The pattern of the device is written on the photoresist by EBL. After developing the photoresist, a 50 nm Cr layer is deposited by electron beam evaporation (EBE) and the Cr hard mask is patterned by the lift-off process. Finally, $Si_3N_4$ nanopillars are etched by RIE and the residual Cr mask is removed by inductively coupled plasma (ICP) etching.

**Supporting Information**

The Supporting Information is available on the website.

**Conflict of Interest**

The authors declare no financial conflicts of interest.

**Ackonwledgements**

The authors are grateful to the partial supports from the Natural Science Foundation of China (Nos. 62075196 and 62105282), Natural Science Foundation of Zhejiang Province (No. LDT23F05014F05), Leading Innovative and Entrepreneur Team Introduction Program of Zhejiang (2021R01001) and the Fundamental Research Funds for the Central Universities.

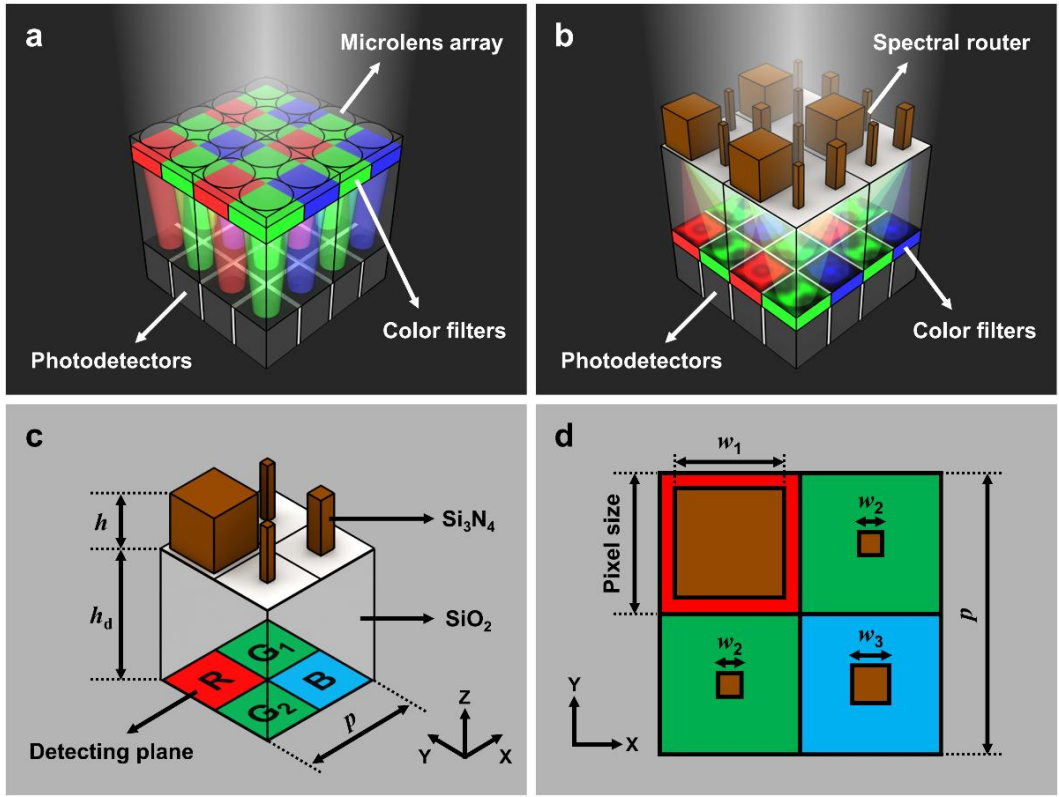

**Figure 1.** Pixelated Bayer spectral router for image sensors. (a) Schematic of a conventional image sensor with the Bayer color filter array. (b) Schematic of an image sensor with the spectral router. (c) Schematic of the spectral router including $Si_3N_4$ nanopillars on the $SiO_2$ layer. $h$ is the height of $Si_3N_4$ nanopillars, $h_d$ is the distance between nanopillars and the detecting plane. (d) Schematic top view of one supercell (2.24 μm × 2.24 μm) of the spectral router, corresponding to one Bayer cell (RGGB) of the image sensor. $w_1$, $w_2$, and $w_3$ represent the widths of nanopillars. $p$ is the size of one supercell.

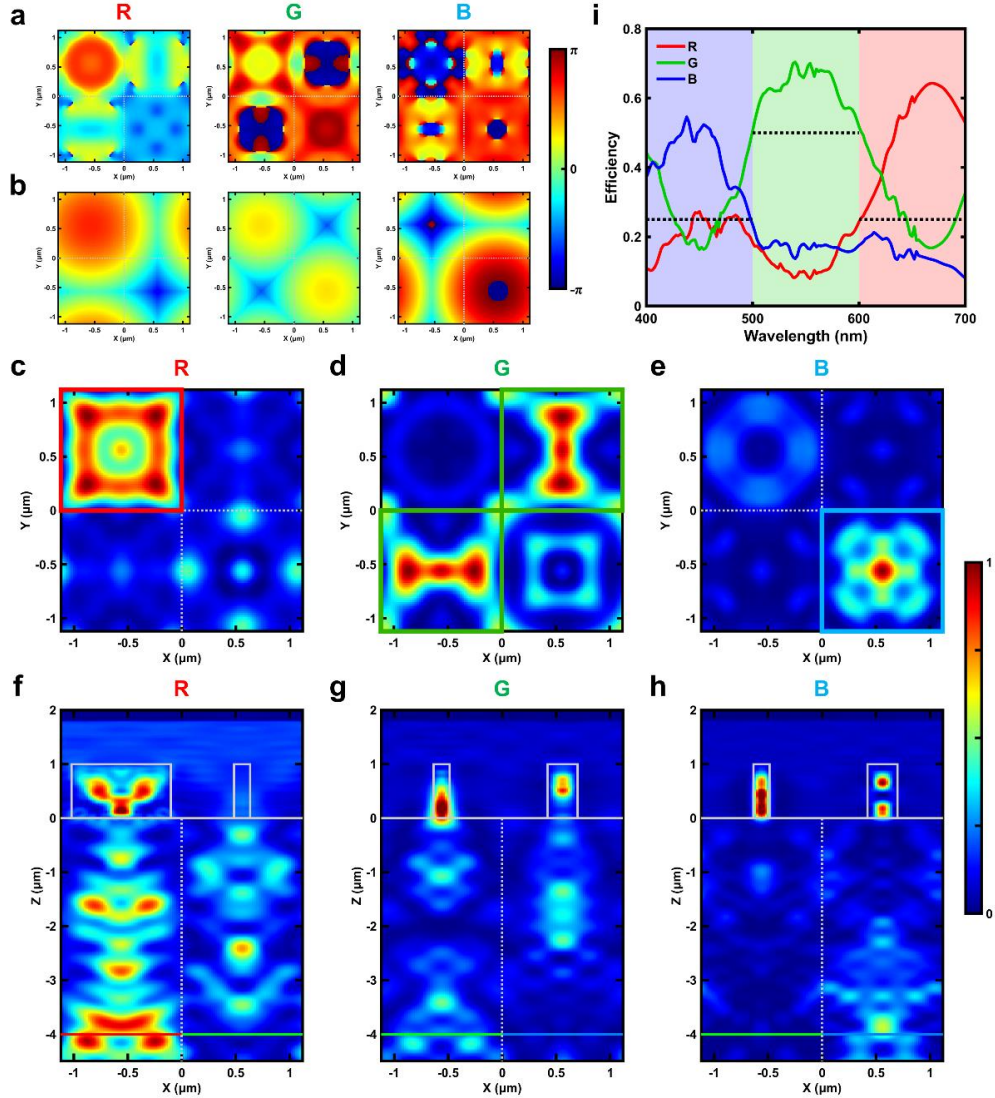

**Figure 2.** Numerical simulation results of the Bayer spectral router. (a) Phase modulation profiles of one supercell of the spectral router at wavelength of 650, 550, and 450 nm. (b) Phase modulation profiles of one supercell of ideal microlens array at wavelength of 650, 550, and 450 nm. (c-e) Simulated power flow density distributions on the detecting plane in one supercell at wavelengths of 650, 550, and 450 nm, respectively. The R, G, and B boxes represent the detecting pixels of corresponding bands. (f-h) Simulated power flow density distributions of the XZ cross section at wavelengths of 650, 550, and 450 nm, respectively. The coordinate system is shown in Figure 1c. Gray rectangular boxes represent $Si_3N_4$ nanopillars. The R, G, and B solid lines at $Z = -4$ μm represent the detecting pixels of corresponding bands. (i) Simulated spectral routing efficiencies of R, G, and B channels. The horizontal dashed lines represent the maximum spectral routing efficiencies of an ideal Bayer color filter array.

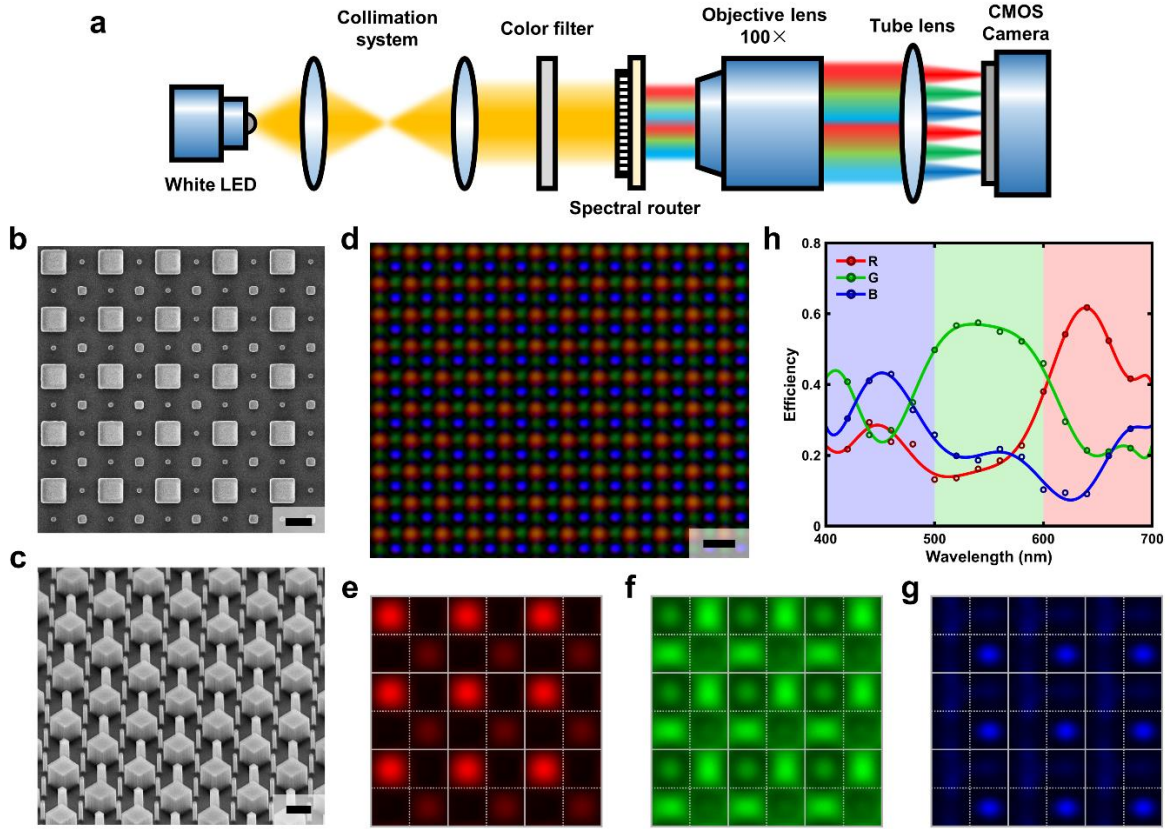

**Figure 3.** Experimental demonstration of the Bayer spectral router. (a) Optical measurement setup for the spectral routing characterization. (b) Top-view, and (c) tilted-view SEM images of the fabricated device. Scale bar: 1 μm. (d) Measured image on the detecting plane of the spectral router under white light illumination. Scale bar: 2.24 μm. (e-g) Measured intensity profiles on the detecting plane of the spectral router under R, G, and B light illumination, respectively. (h) Measured spectral routing efficiencies of R, G, and B channels.

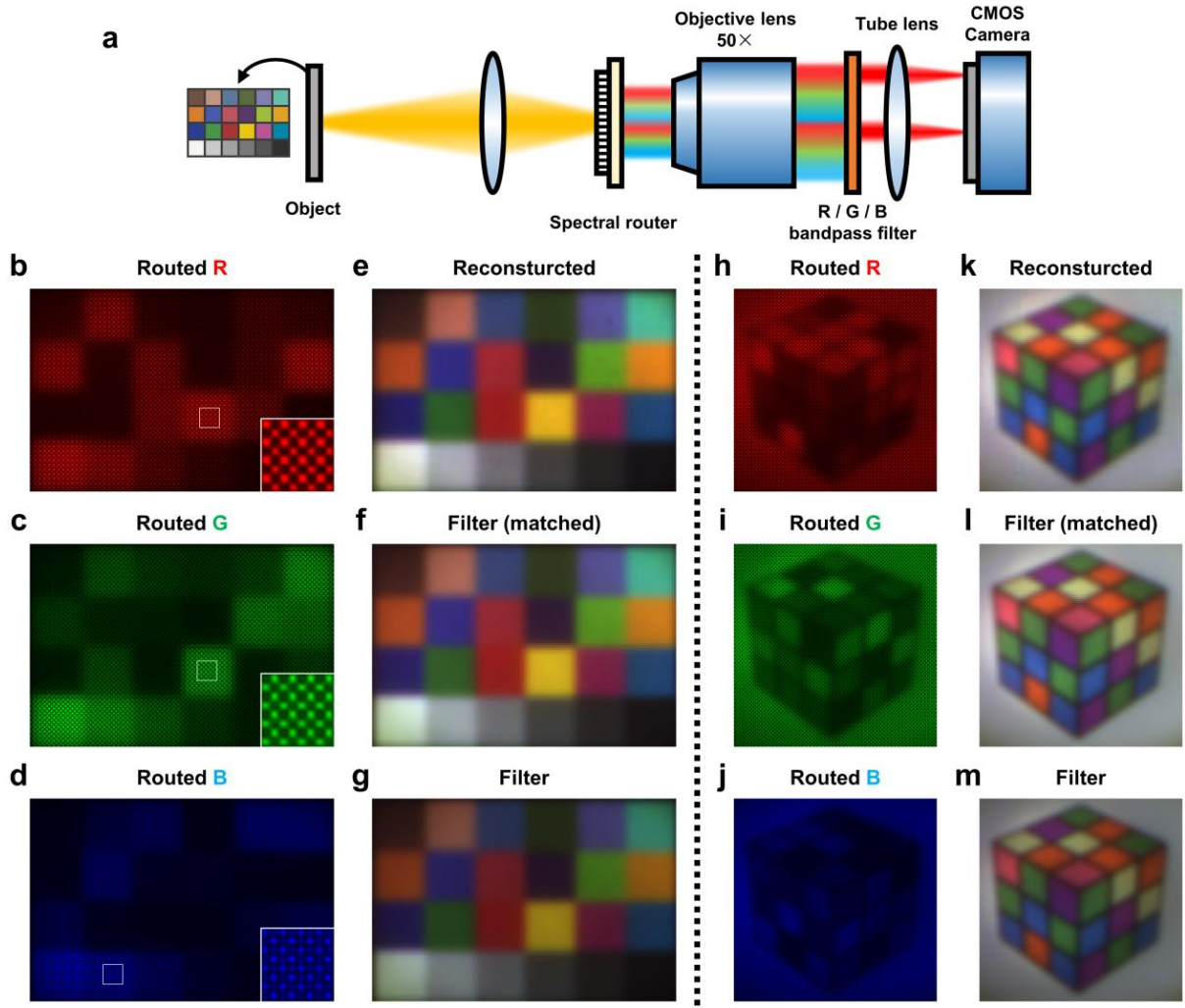

**Figure 4.** Color imaging with the Bayer spectral router. (a) Experimental setup for color imaging utilizing the spectral router. R, G, and B color filters are employed successively to mimic the Bayer color filter array on the detecting plane to eliminate crosstalk. (b-d) Measured intensity profiles of R, G, and B channels, respectively, on the detecting plane after the image of the ColorChecker is routed. The insets show the enlarged images of areas in the white boxes. (e) Reconstructed color image of the ColorChecker. (f,g) Reference color image obtained using only color filters without the spectral router. Intensity of (f) is matched to that of (e). (h-m) Analogous experimental results to (b)-(g), reconstructing the color image of the Rubik's Cube.

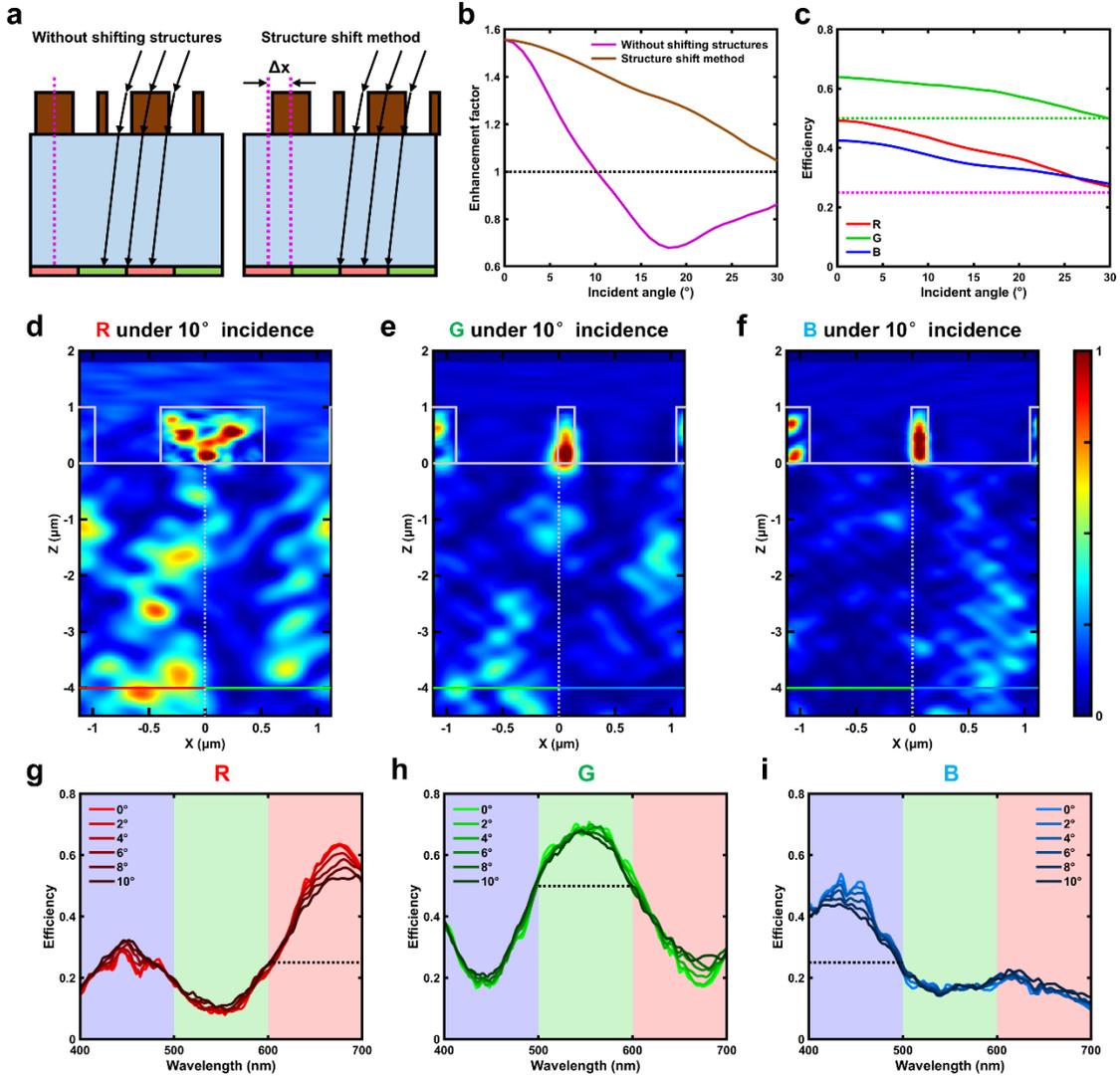

**Figure 5.** Structure shift method for improving incident angle tolerance. (a) Schematics of the light propagation trajectory without and with the structure shift method. $\Delta x$ represents the shift length of structures. (b) Efficiency enhancement factor as a function of the incident angle with and without shifting structures. (c) Average spectral routing efficiencies of R, G, and B channels as a function of the incident angle when using the structure shift method. (d-f) Simulated power flow density distributions of the XZ cross section at wavelengths of 650, 550, and 450 nm, respectively, under 10° incidence when using the structure shift method. Gray rectangular boxes represent $Si_3N_4$ nanopillars. The R, G, and B solid lines at Z = -4 µm represent the detecting pixels of corresponding bands. (g-i) Simulated spectral routing efficiencies of R, G, and B channels, respectively, under different incident angles when using the structure shift method.

# Supporting Information

# Pixelated Bayer Spectral Router Based on Sparse Meta-atom Array

**Yifan Shao, Rui Chen, Yubo Wang, Shuhan Guo, Junjie Zhan, Pankaj K. Choudhury, and Yungui Ma\***

State Key Lab of Modern Optical Instrumentation, Centre for Optical and Electromagnetic Research, College of Optical Science and Engineering; International Research Center (Haining) for Advanced Photonics, Zhejiang University, Hangzhou, 310058, China

**Corresponding author's E-mail:** yungui@zju.edu.cn

## Section S1. Optical constants of used materials

Figure S1 plots the optical constants of $Si_3N_4$ and $SiO_2$. The refractive index and extinction coefficient of $Si_3N_4$ are measured by an ellipsometer. The refractive index of $SiO_2$ is obtained from the data measured by Malitson in 1965.

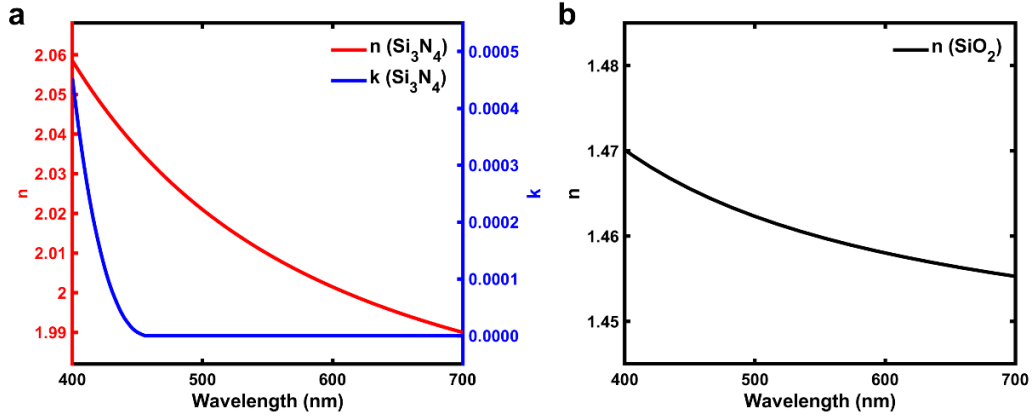

**Figure S1.** (a) The refractive index and extinction coefficient of $Si_3N_4$. (b) The refractive index of $SiO_2$.

## Section S2. Optimization process of the Bayer spectral router

The spectral routing efficiencies of R, G, and B channels as a function of wavelength are defined as

$$T_R(\lambda) = \frac{\int_R I(x, y, \lambda)\, dxdy}{\int_{R+G_1+G_2+B} I_i(x, y, \lambda)\, dxdy} \tag{S1}$$

$$T_G(\lambda) = \frac{\int_{G_1+G_2} I(x, y, \lambda)\, dxdy}{\int_{R+G_1+G_2+B} I_i(x, y, \lambda)\, dxdy} \tag{S2}$$

$$T_B(\lambda) = \frac{\int_B I(x, y, \lambda)\, dxdy}{\int_{R+G_1+G_2+B} I_i(x, y, \lambda)\, dxdy} \tag{S3}$$

respectively, where $I(x, y, \lambda)$ represents the intensity distribution on the detecting plane at the wavelength of $\lambda$, $I_i(x, y, \lambda)$ represents the intensity distribution of incident light at the wavelength of $\lambda$, the subscripts of integral symbols indicate the pixel regions where the integral is performed. Figure S2 (b) depicts the layout of R, $G_1$, $G_2$, and B pixels in the Bayer pattern.

The average spectral routing efficiencies of R, G, and B bands are defined as

$$\text{Eff}_R = \frac{1}{\Delta\lambda_R} \int_{600\text{ nm}}^{700\text{ nm}} T_R(\lambda)\, d\lambda \tag{S4}$$

$$\text{Eff}_G = \frac{1}{\Delta\lambda_G} \int_{500\text{ nm}}^{600\text{ nm}} T_G(\lambda)\, d\lambda \tag{S5}$$

$$\text{Eff}_B = \frac{1}{\Delta\lambda_B} \int_{400\ \text{nm}}^{500\ \text{nm}} T_B(\lambda)\, d\lambda \tag{S6}$$

respectively, where $\Delta\lambda_R = \Delta\lambda_G = \Delta\lambda_B = 100$ nm are bandwidths of R, G, and B bands.

Particle swarm optimization (PSO) algorithm is a robust stochastic evolutionary computation technique for finding the optimal solution in complex search spaces through the interaction of individuals in a population of particles,[1,2] which has long been applied in electromagnetics.[3] In this work, we combine the PSO algorithm with FDTD simulation to find the optimal structure parameters for the spectral router. The figure of merit (FOM, i.e. fitness function) is set as

$$\text{Eff} = \text{Eff}_R + \text{Eff}_G + \text{Eff}_B \tag{S7}$$

which is the efficiency enhancement factor compared with the maximum efficiency of traditional Bayer image sensors with ideal color filters (100% transmittance in the passband). The FOM intends to obtain the highest possible spectral routing efficiency.

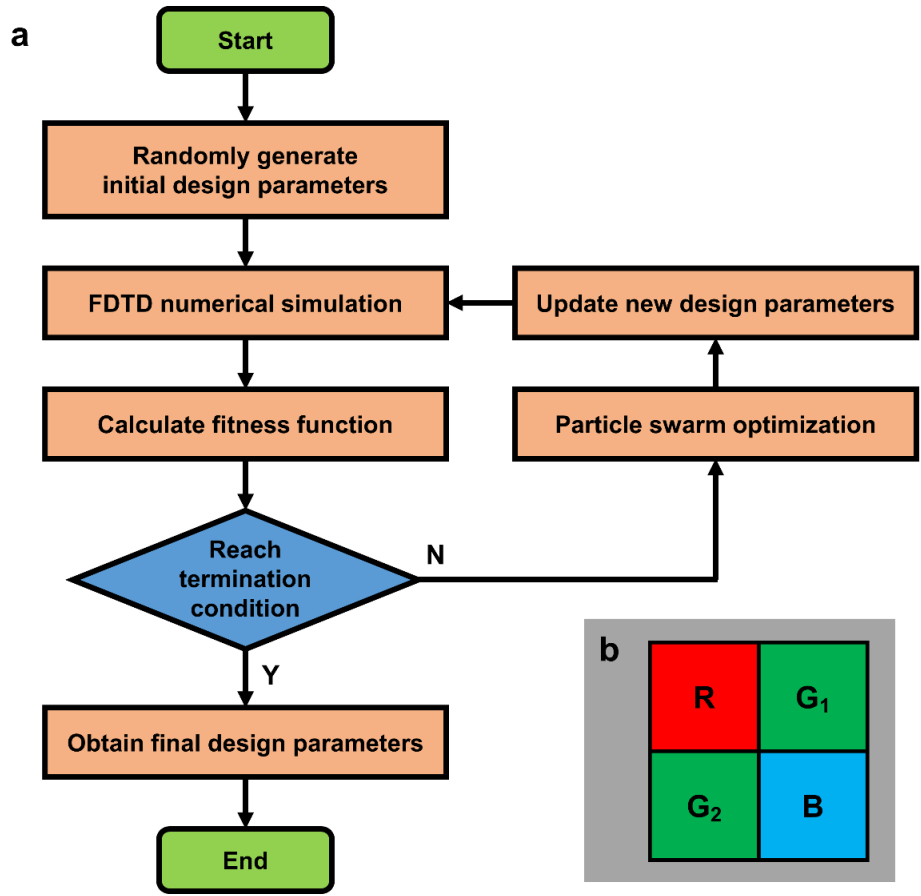

**Figure S2.** (a) Optimization flow diagram of the pixelated spectral router. (b) Schematic top view of one supercell of the spectral router, corresponding to one Bayer cell (RGGB) of the image sensor.

The number of solutions (particles) in each generation is set to 20, and the maximum generation number is set to 1000. First, the initial design parameters are randomly generated. Then, as the optimization flow diagram in Figure S2 (a), in each iteration, numerical simulation of the spectral router is performed by Lumerical FDTD Solutions, followed by the calculation of FOM (fitness function). The criterion for optimization convergence is that the increase of FOM is less than $10^{-6}$ in 100 consecutive generations (i.e., 2000 solutions). The optimization is repeated until the convergence condition is met or the maximum number of generation is reached. Ultimately, the final design parameters are established for the fabrication of the spectral router.

**Section S3. The focusing effect for R, G, and B color under limited aperture illumination**

In order to demonstrate the working principle of sparse meta-atoms as dispersion-engineered micro-metalenses, Figure 2a in the main text shows the phase modulation profiles of one supercell for R, G, and B light, which are similar to those of ideal microlens array that meets the following phase distribution

$$\varphi_\lambda(x, y) = -\frac{2\pi}{\lambda}\left(\sqrt{f^2 + (x - x_\lambda)^2 + (y - y_\lambda)^2} - f\right) + C_\lambda \quad \text{(S8)}$$

especially for R and B light. In Equation S8, $f$ is the focal length, $x$ and $y$ are coordinates of X-axis and Y-axis, $(x_\lambda, y_\lambda)$ is the focus coordinate at the wavelength of $\lambda$, and $C_\lambda$ is bias of the phase at the wavelength of $\lambda$. The phase modulation profile of G light deviates more from the ideal value, resulting in an inferior focusing effect compared to that of R and B light.

Figure S3 demonstrates the simulated focusing effect for R, G, and B color under limited aperture illumination, where the incident light only cover nine meta-atoms (regions enclosed by the white boxes in Figure S3 (a-c)). For R, $G_2$, and B pixels, respectively, corresponding nanopillars are arranged only on the central pixel and the eight surrounding pixels. R, G, and B arrows in Figure S3 (a-c) indicate the focusing direction of light. As shown in Figure S3 (d-i), the sparse meta-atoms function as a microlens, well focusing light towards the central pixel, thus making the central pixel collect energy that would otherwise be absorbed by color filters of the neighboring pixels of different colors. Note that for G light, the surrounding eight meta-atoms not only focus light towards the central $G_2$ pixel, but also route a portion of light to the four $G_1$ pixels in the four corners of the aperture. However, the four $G_1$ pixels in the four corners lack the light routed from adjacent pixels outside the white box in Figure S3 (b), which is indicated by the laurel-green dashed arrows in Figure S3 (b). Therefore, these four $G_1$ corners do not form perfect focal spots in this simulation. According to the above analysis, the contribution to the focusing effect primarily comes from the central meta-atoms and the neighboring meta-atoms on the pixels of colors different from the central pixel, indicating that the spatial crosstalk of the same color channel between supercells is small.

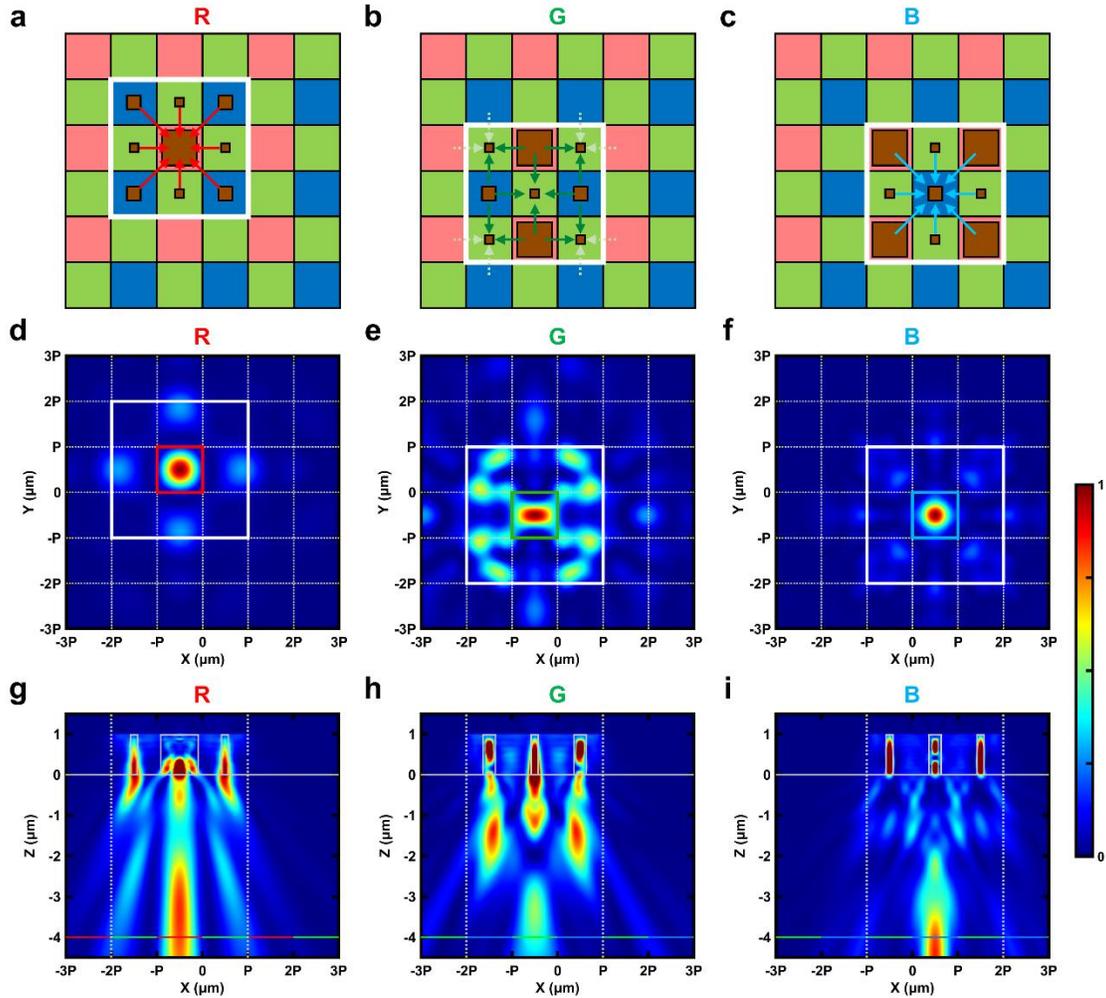

**Figure S3.** The focusing effect for R, G, and B color under limited aperture illumination. (a-c) Schematics: For R, $G_2$, and B pixels, respectively, corresponding nanopillars are arranged only within the adjacent pixels. The white boxes indicate the aperture ranges of the incident light, covering nine meta-atoms. R, G, and B arrows indicate the focusing direction of light. (d-f) Power flow density distributions on the detecting plane at wavelengths of 650, 550, and 450 nm, respectively. The white boxes indicate the aperture ranges of incident light, covering nine pixels. (g-i) Power flow density distributions of the XZ cross section at wavelengths of 650, 550 and 450 nm, respectively. Gray rectangular boxes represent $Si_3N_4$ nanopillars. The R, G, and B solid lines

at Z = -4 µm represent the detecting pixels of corresponding bands. The vertical dashed lines indicate the aperture ranges of incident light.

## Section S4. Fabrication of the Bayer spectral router

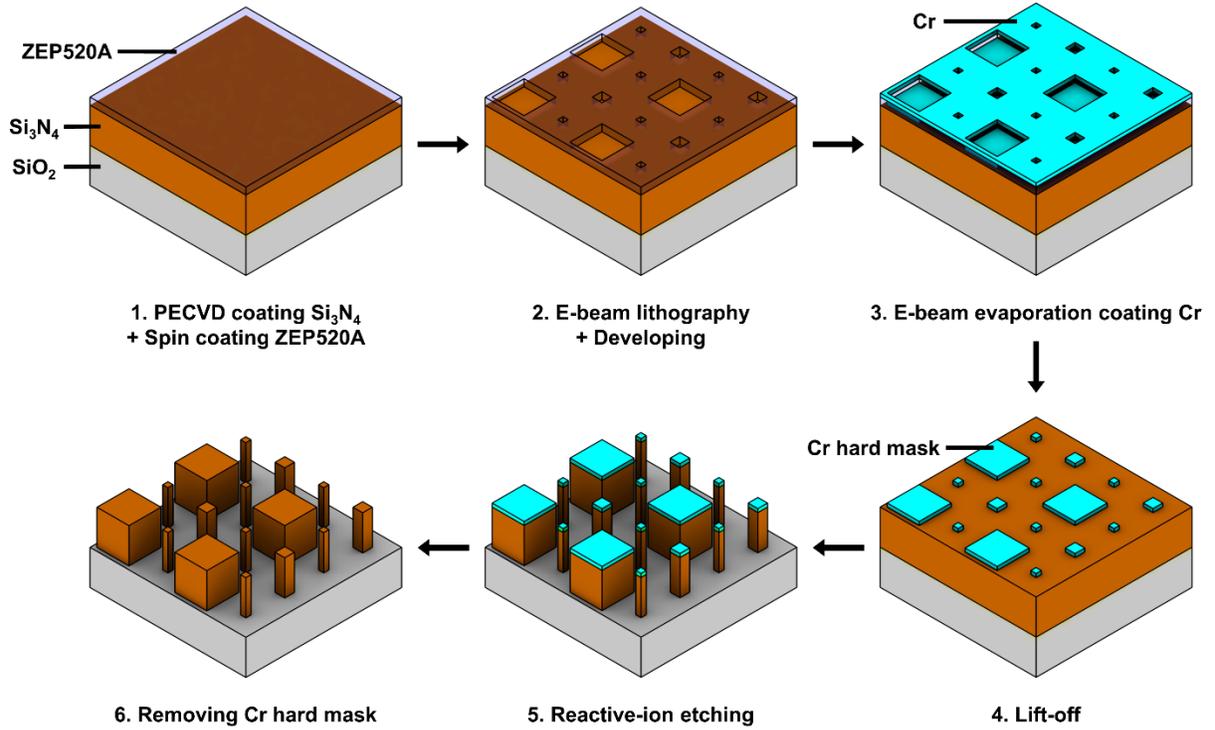

**Figure S4.** Flow chart depicting the fabrication steps of the spectral router.

## Section S5. Experimental characterizations of the Bayer spectral router

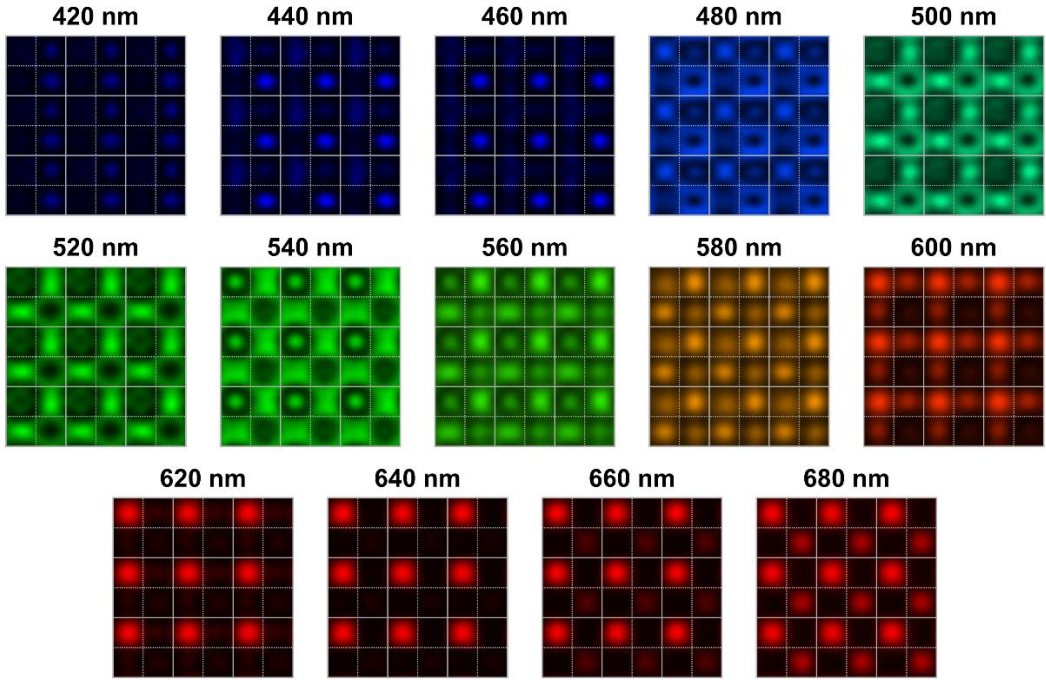

**Figure S5.** Measured intensity profiles on the detecting plane of the spectral router at different wavelengths.

## Section S6. Spectra of RGB bandpass color filters

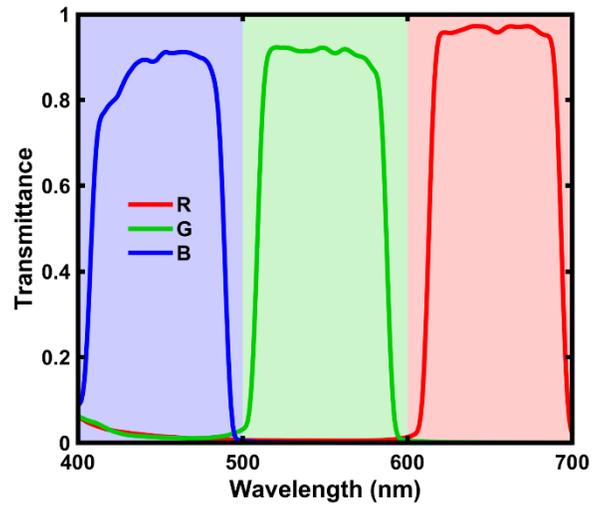

**Figure S6.** Spectra of R, G, and B bandpass color filters used in the experiment of color imaging.

**Section S7. Color imaging with the Bayer spectral router (ColorChecker)**

The experimental setup for color imaging is shown in Figure 4a of the main text. The object (ColorChecker) is placed 665 mm away from the imaging system and is imaged onto the spectral router. Figure S7 (a-c) present the measured intensity profiles on the detecting plane after the image is routed, which are obtained by inserting R, G, and B bandpass color filters, respectively. Then, the mosaic images of R, G, and B channels can be obtained directly according to the calibrated spectral routing efficiency. Figure S7 (d-f) present the reconstructed images of R, G, and B channels, respectively, after demosaicing the images through bilinear interpolation. Figure S7 (h-k) present the reference images of R, G, and B channels obtained using the same experimental setup without the spectral router, which are also reconstructed by color correction and demosaicing. Figure S7 (d-f) are in agreement with Figure S7 (h-j). Figure S7 (g) and (k) are reconstructed color images with and without the spectral router after performing white balance, also showing a good agreement.

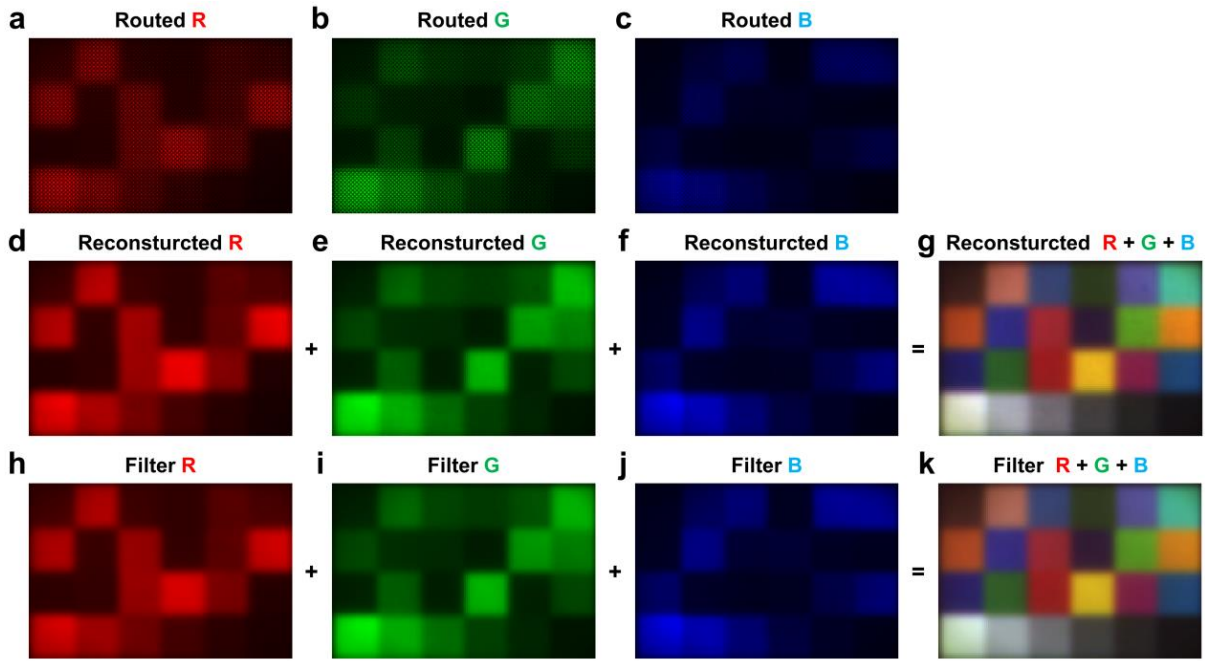

**Figure S7.** Color imaging of the ColorChecker with the spectral router. (a-c) Measured intensity profiles of R, G, and B channels, respectively, on the detecting plane after the image is routed. (d-f) Reconstructed images of R, G, and B channels, respectively. (g) Reconstructed color image obtained from panels (d)-(f). (h-j) Reference images of R, G, and B channels obtained using only color filters without the spectral router. (k) Reference color image obtained from panels (h)-(j).

**Section S8. Color imaging with the Bayer spectral router (Rubik's Cube)**

The object (Rubik's Cube) is placed 540 mm away from the imaging system. The other experimental process and results are all analogous to Section S7.

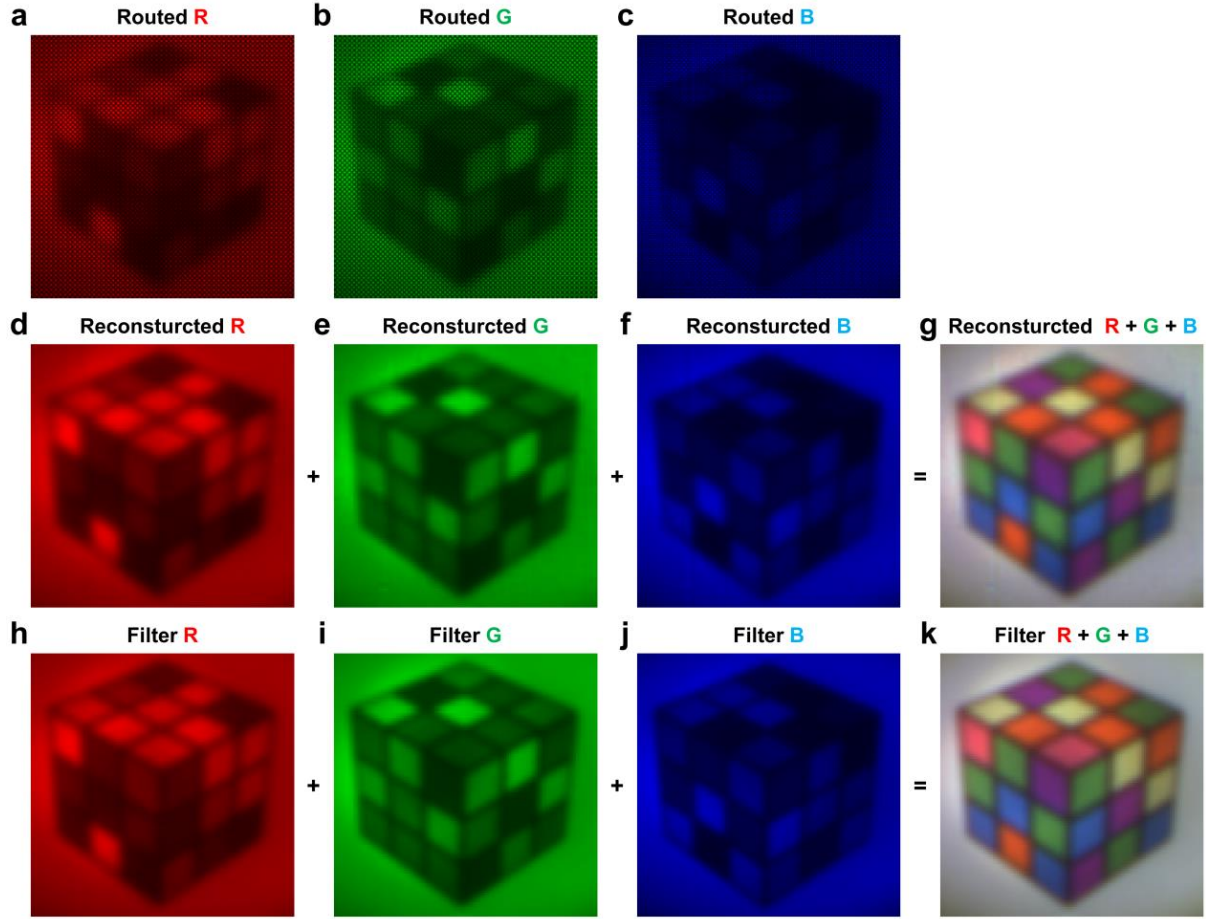

**Figure S8.** Color imaging of the Rubik's Cube with the spectral router. (a-c) Measured intensity profiles of R, G, and B channels, respectively, on the detecting plane after the image is routed. (d-f) Reconstructed images of R, G, and B channels, respectively. (g) Reconstructed color image obtained from panels (d)-(f). (h-j) Reference images of R, G, and B channels obtained using only color filters without the spectral router. (k) Reference color image obtained from panels (h)-(j).

S9

## Section S9. Efficiency enhancement for color imaging with the Bayer spectral router

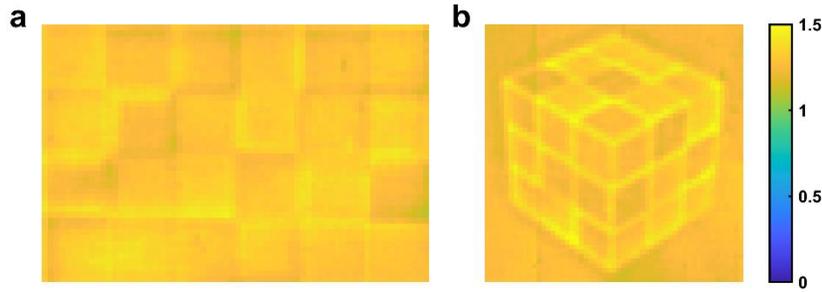

**Figure S9.** Efficiency enhancement factor profile for color imaging with the Bayer spectral router. (a) ColorChecker, (b) Rubik's Cube.

## Section S10. Polarization insensitivity of the Bayer spectral router

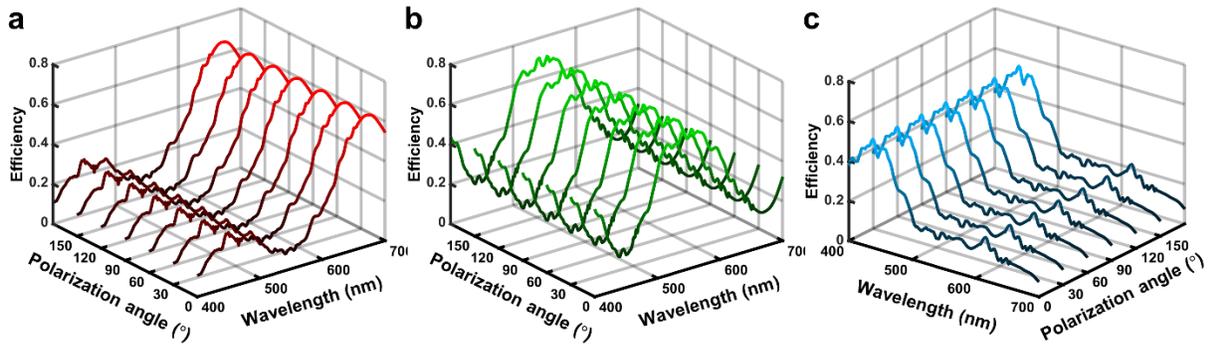

**Figure S10.** Spectral routing efficiencies of (a) R, (b) G, and (c) B channels with different polarization angles.

## Section S11. Incident angle tolerance and the specially optimized design under oblique incidence

Figure S11 (a-c) plot simulated spectral routing efficiencies of R, G, and B channels, respectively, under different incident angles. The average efficiency of R, G, and B channels remains higher than 33% up to a 10° incident angle. Figure S11 (d-f) illustrate the simulated power flow density distributions of the XZ cross section under 10° incidence without shifting structures. Some light intended to be routed to $R/G_1/G_2/B$ pixels is misrouted to other pixels respectively due to the off-axis illumination, which causes severe crosstalk between color channels.

As shown in Figure 5 of the main text, the device efficiency under oblique incidence can be improved by shifting nanostructures according to the chief ray angle corresponding to the pixel position. The maximum acceptable chief ray angle of the spectral router can be expanded to over 30° through the structure shift method. Moreover, the spectral routing efficiency under the illumination of large incident angles can even be further improved by specially optimizing the nanostructures while keeping the height of nanopillars and the position of detecting plane unchanged ($h$ = 998 nm, $h_d$ = 4μm). Figure S11 (h) plots the spectral routing efficiency under 30° incidence with specially optimized parameters ($w_1$ = 792 nm, $w_2$ = 205 nm, $w_3$ = 150 nm, $\Delta x$ = -200 nm).

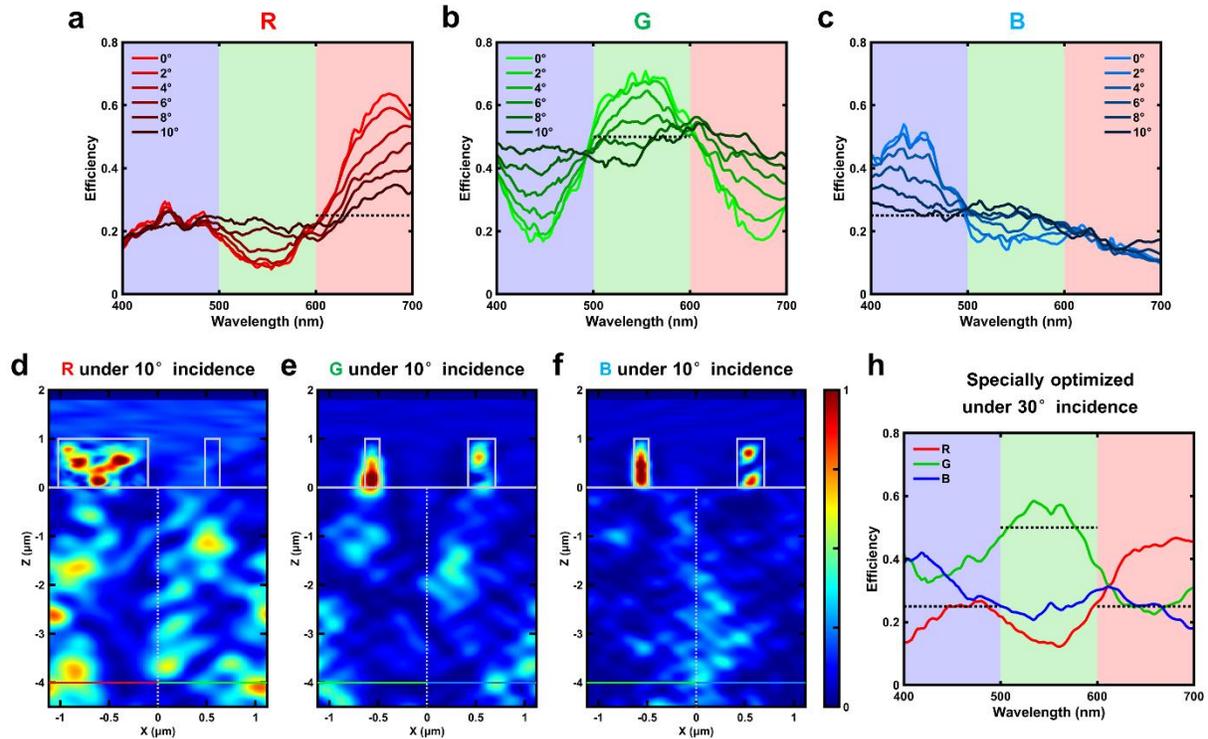

**Figure S11.** (a-c) Simulated spectral routing efficiencies of R, G, and B channels, respectively, under different incident angles. (d-f) Simulated power flow density distributions of the XZ cross section at wavelengths of 650, 550, and 450 nm, respectively, under 10° incidence without shifting structures. Gray rectangular boxes represent $Si_3N_4$ nanopillars. The R, G, and B solid lines at Z = -4 μm represent the detecting pixels of corresponding bands. (h) Simulated spectral routing efficiencies of R, G, and B channels under 30° incidence with specially optimized structures.

**Section S12. Comparison between this work and previous spectral router**

Some spectral routers were designed via single-layer code-like or freeform metasurfaces.[4-7] Figure S12 (a) illustrates the layout of a Bayer spectral router based on code-like metasurfaces consisting of many fine nanostructures, which is like a chessboard.[4] Figure S12 (b) illustrates the schematic top view of the Bayer spectral router designed in our work. The structural layout of our spectral router is very simple, with a relatively large feature size (150 nm) and all gaps between nanopillars larger than 500 nm, which is very promising for massive production. As shown in Figure S12 (c-e), the spectral router based on code-like metasurfaces exhibits high performance in the R and B channels, whereas the spectral routing efficiency of the G channel is about on par with that of the traditional color filter scheme. In this study, our spectral router notably improves the spectral routing efficiency of the G channel while maintaining the high performance of the R and B channels.

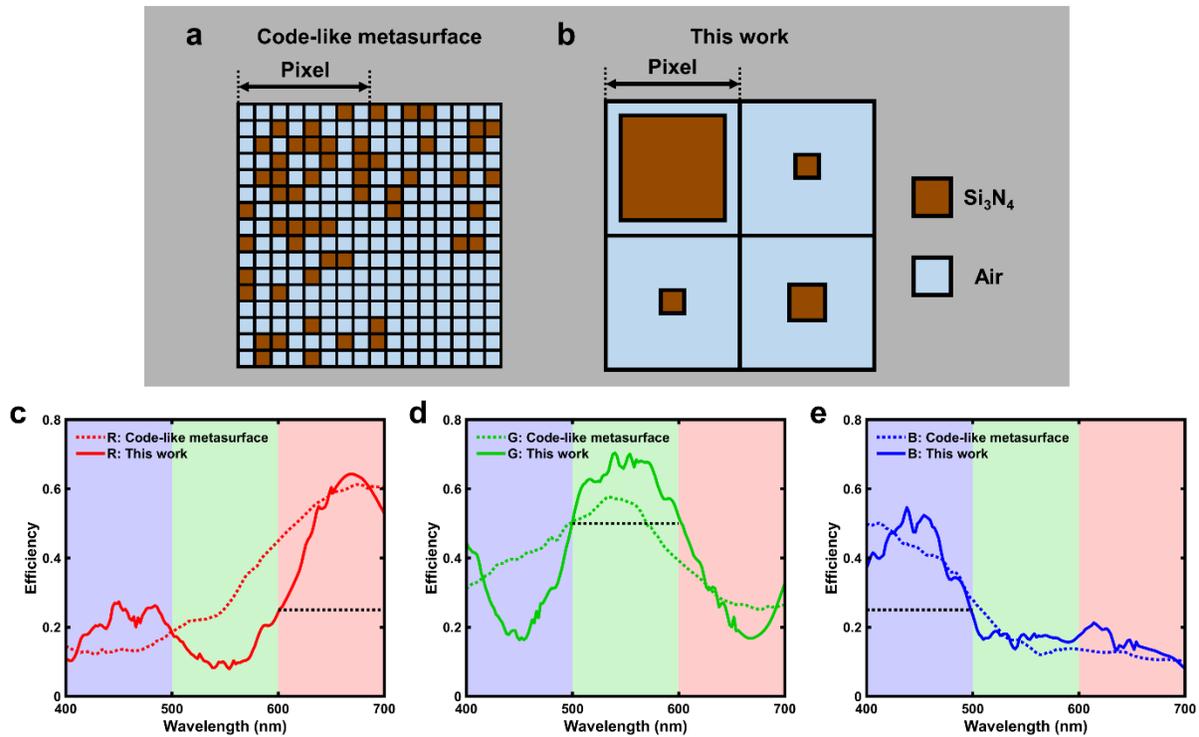

**Figure S12.** (a) Structural layout of the spectral router based on code-like metasurfaces. (b) Structural layout of the spectral router in this work. (c-e) Comparison between efficiencies of spectral routers based on code-like metasurfaces and this work: spectral routing efficiencies of R, G, and B channels, respectively.

To intuitively demonstrate the advantages of our router, Table S1 compares the feature sizes, gaps between nanostructures, pixel sizes, and efficiency enhancement factors of this work with those of previous spectral routers. To avoid the influence of device fabrication errors and experimental measurement errors for a fair comparison, the efficiency enhancement factors are uniformly compared by the simulation design results. Compared to other spectral routers, our design can achieve similar or even higher efficiency while having the simpler structure layout, larger feature size and gaps between nanostructures. It is of great significance for massive production in practical industrial applications.

**Table S1.** Comparison between single-layer spectral routers

| Reference | Scheme | Feature size | Gaps between nanosturctures | Pixel size | Efficiency enhancement factor (R + G + B) |
|---|---|---|---|---|---|
| Ref. [12] | Blazed grating | 125 nm | 190 nm | 1.43 μm | (40-50%) × 3 |
| Ref. [20] | Micro-metalens array | ~80 nm | ~100 nm | 1.6 μm | 65% + 50% + 25% = 1.4 |
| Ref. [25] | Micro-metalens array | 50 nm | ~100 nm | 25 μm | 15.9% + 65.42% + 38.33% = 1.197* |
| Ref. [26] | Code-like metasurface | 125 nm | 125 nm | 1 μm | ~56.2% + 51.4% + 41.8% = 1.494 |
| Ref. [28] | Code-like metasurface | 100 nm | 100 nm | 1.1 μm | ~ 49% + 49% + 42% = 1.4 |
| Ref. [29] | Freeform metasurface | 50 nm | ~50 nm | 0.6 μm | ~52%+52%+56% = 1.6<br>60%+57%+65% = 1.82* |
| This work | Sparse meta-atoms | **150 nm** | **>500 nm** | 1.12 μm | **51.13% + 62.91% + 42.57% = 1.566**<br>**64.30% + 70.43% + 54.62% = 1.894*** |

**Note:** The reference numbers in Table S1 are corresponding to the reference numbers in the main text. All the efficiency enhancement factors are simulation design results to avoid the influence of device fabrication errors and experimental measurement errors. Asterisk * means the peak spectral routing efficiencies. The others are broadband average spectral routing efficiencies.

## Section S13. Design of Bayer spectral routers for other pixel sizes

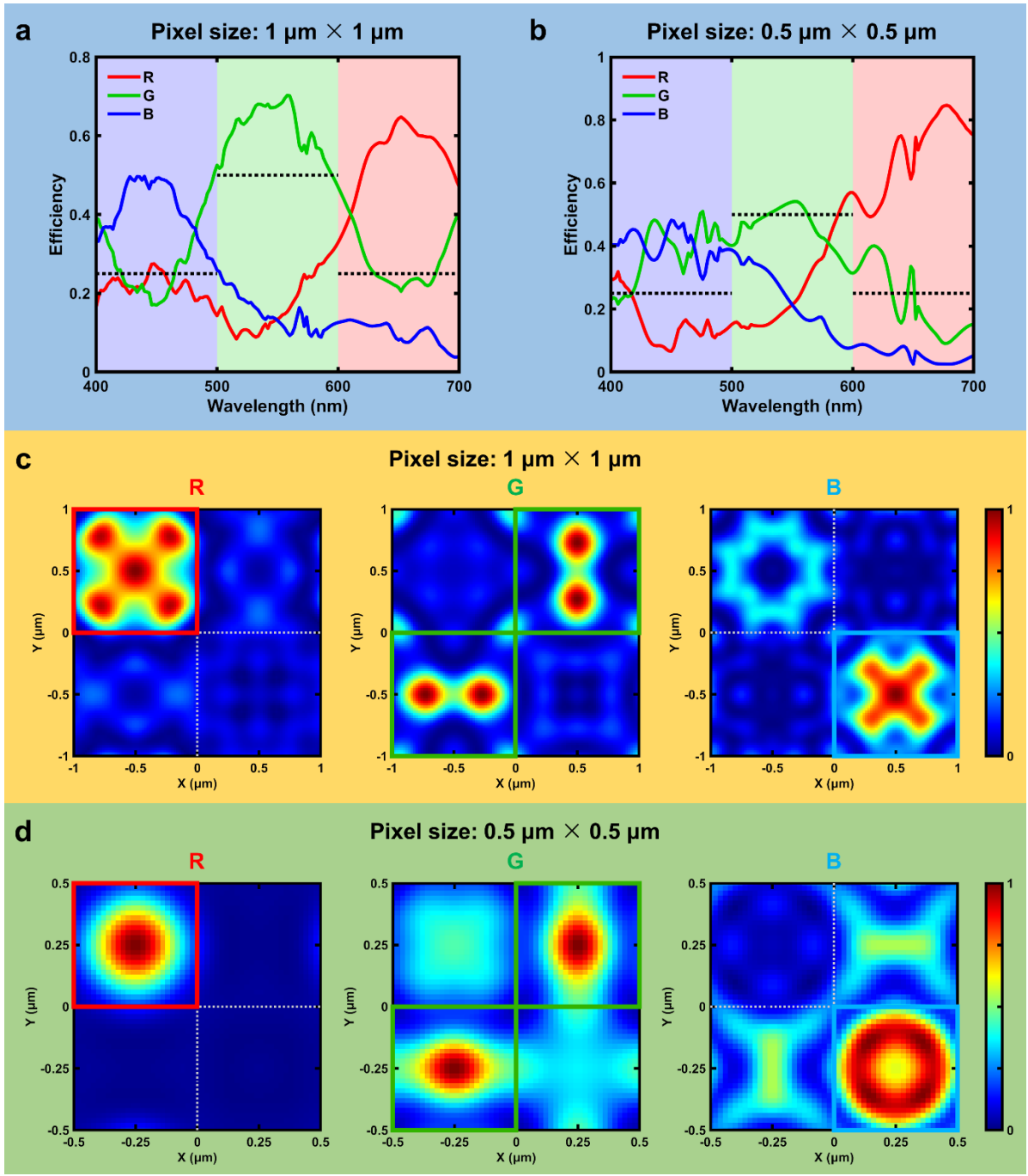

**Figure S13.** Simulated spectral routing efficiencies of R, G, and B channels of the spectral router with (a) 1 μm × 1 μm pixel size, and (b) 0.5 μm × 0.5 μm pixel size. Simulated power flow density distributions on the detecting plane in one supercell of the spectral router with (c) 1 μm × 1 μm pixel size, and (d) 0.5 μm × 0.5 μm pixel size at wavelengths of 650, 550, and 450 nm.

The supercell of the spectral router in the main text has a period of 2.24 μm ×2.24 μm, corresponding to one Bayer cell (RGGB) with the pixel size of 1.12 μm × 1.12 μm. In fact, our method of designing spectral routers based on sparse meta-atom array can be flexibly applicable to other pixel sizes. Here, we demonstrate Bayer spectral routers that are matched to pixel sizes of 1 μm × 1 μm and 0.5 μm × 0.5 μm, which are designed by the same structural layout in the main text.

(1) Bayer spectral router matched to the pixel size of 1 μm × 1 μm

Design parameters: period = 2 μm × 2 μm, $w_1$ = 928 nm, $w_2$ = 150 nm, $w_3$ = 276 nm, $h$ = 986 nm, and $h_d$ = 3 μm. Figure S13 (a) plots the simulated spectral routing efficiencies of R, G, and B channels. Figure S13 (c) shows the simulated power flow density distributions on the detecting plane.

(2) Bayer spectral router matched to the pixel size of 0.5 μm × 0.5 μm

Design parameters: period = 1 μm × 1 μm, $w_1$ = 410 nm, $w_2$ = 174 nm, $w_3$ = 222 nm, $h$ = 720 nm, and $h_d$ = 2.48 μm. Figure S13 (b) plots the simulated spectral routing efficiencies of R, G, and B channels. Figure S13 (d) shows the simulated power flow density distributions on the detecting plane.

**Section S14. Integration between the spectral router, color filter array and photodetector array**

Figure S14 (a) illustrates the schematic side view of the image sensor that integrates the spectral router, color filters and silicon photodetectors. The thickness of color filters is set to 500 nm in the simulation [Z between 0 and 0.5 μm in Figure S14 (f-h)]. The refractive index of color filters is set to 1.5, and the extinction coefficient is set to 0.3 in the stopbands of color filters. The $Si_3N_4$ anti-reflection layer with a thickness of 70 nm can increase the transmittance. Figure S14 (c-e) illustrate the power flow density distributions on the photodetectors at wavelengths of 650, 550, and 450 nm, respectively. Figure S14 (f-h) illustrate the power flow density distributions in the propagation direction at wavelengths of 650, 550, and 450 nm, respectively. It is obvious that the crosstalk between different channels is significantly reduced by the color filter array. Figure S14 (b) compares the simulated efficiencies between two architectures: (i) integrating the spectral router with color filters, and (ii) integrating the microlens array with color filters. It obviously reflects the significant role of spectral router in the efficiency enhancement.

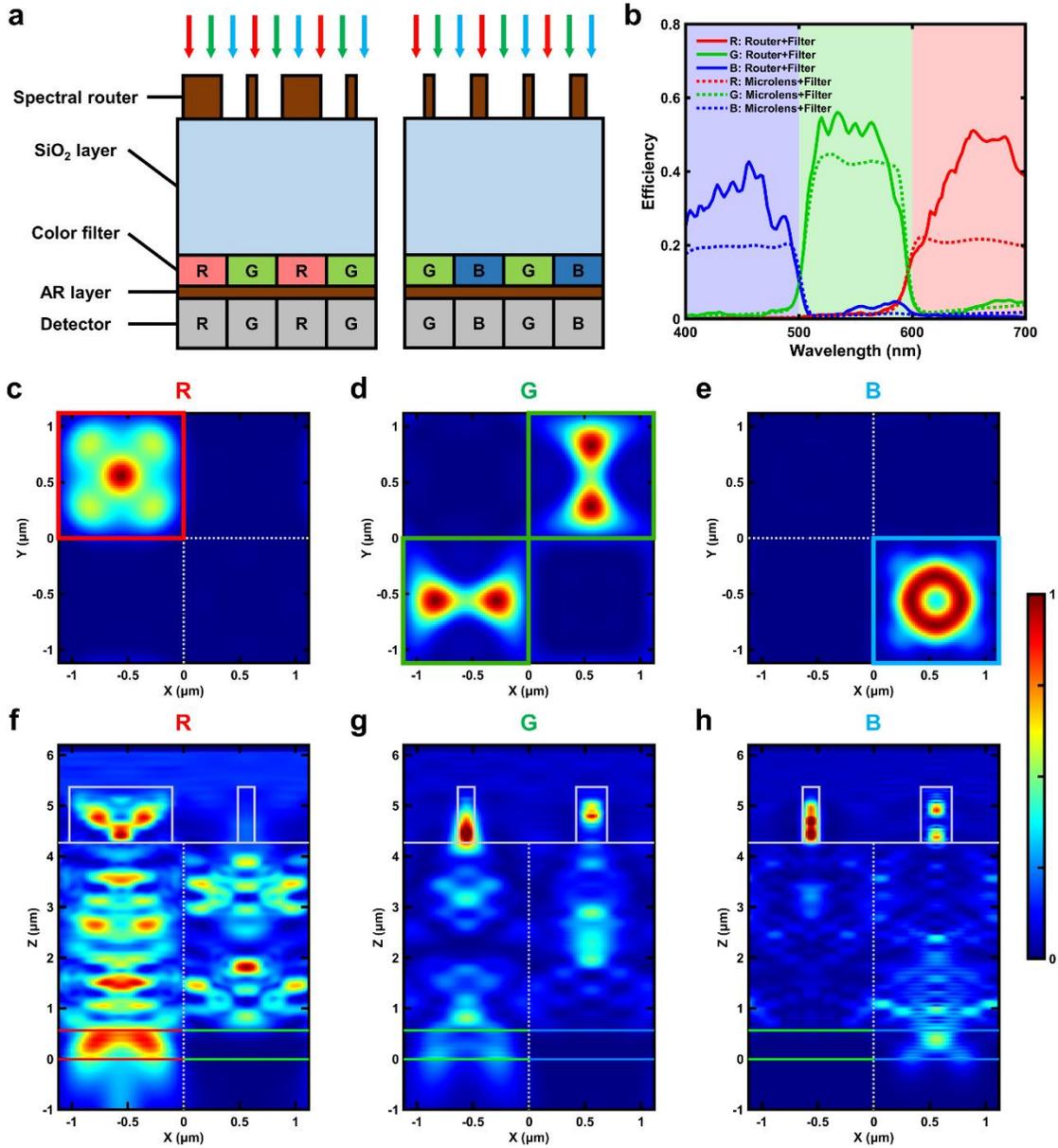

**Figure S14.** (a) Schematic side view of the image sensor that integrates the spectral router, color filters, and photodetectors. (b) Comparison between efficiencies of combining the spectral router and color filters, and combining the microlens array and color filters (conventional image sensor). (c-e) Simulated power flow density distributions on the photodetectors after light passes through the spectral router and color filters at wavelengths of 650, 550, and 450 nm, respectively. (f-h) Simulated power flow density distributions of the XZ cross section at wavelengths of 650, 550, and 450 nm, respectively. Gray rectangular boxes represent $Si_3N_4$ nanopillars. Regions between two R/G/B solid lines represent corresponding color filters which can eliminate crosstalk.

**References of Supporting Information**